\documentclass{PoS}

\def \inte {$INTEGRAL$}

\def \sw {$Swift$}

\def \hcm {\hbox {\ifmmode $ atom cm$^{-2}\else atom cm$^{-2}$\fi}}

\def \aap {A\&A}

\def \mnras {MNRAS}

\title{Supergiant Fast X--ray Transients}

\ShortTitle{Supergiant Fast X--ray Transients }

\author{\speaker{L.\ Sidoli} \thanks{Invited talk}\\
        INAF, Istituto di Astrofisica Spaziale e Fisica Cosmica, \\
         Via E.\ Bassini 15,   I-20133 Milano,  Italy\\
        E-mail: \email{sidoli@iasf-milano.inaf.it}}

\abstract{The phenomenology of a subclass of High Mass X--ray Binaries 
hosting a blue supergiant companion, the so-called
Supergiant Fast X--ray Transients (SFXTs), is reviewed. 
Their number is growing, mainly thanks to the discoveries
performed by the \inte\ satellite, then followed by soft X--rays observations (both aimed
at refining the source position and at monitoring the source behavior) 
leading to the optical identification of the blue supergiant nature of the donor star.
Their defining properties are a transient X--ray activity consisting of sporadic, fast and bright flares, 
together with the association with an OB supergiant.
The SFXTs outbursts are characterized by a duration of a few days, composed by several flares (each 
with a variable duration between a few minutes and a few hours), 
reaching 10$^{36}$--10$^{37}$~erg~s$^{-1}$.
The quiescence is at a luminosity of 10$^{32}$~erg~s$^{-1}$, while 
their more frequent state consists of an intermediate X--ray emission of 
10$^{33}$--10$^{34}$~erg~s$^{-1}$ (1--10~keV).
Only the brightest flares  are detected by \inte\ ($>$17~keV) during short
pointings, with no detected persistent emission.
The physical mechanism driving the short outbursts is still debated, although a huge amount of
observational data have been collected, particularly aimed at searching 
for pulse and orbital periodicities.
About a half of the members of the class displays X--ray pulsations, indicative of the neutron star
nature of the compact object. In other SFXTs a black hole cannot be excluded,
although the X--ray spectrum in outburst resembles that shown by  accreting X--ray pulsars.
Since their transient activity is brief, although 
recurrent with different timescales from source
to source, they could represent a numerous class 
of Galactic massive X--ray binaries, 
remained hidden up to now.
}

\FullConference{25th Texas Symposium on Relativistic Astrophysics - TEXAS 2010\\
		December 06-10, 2010\\
		Heidelberg, Germany}

\begin{document}

\section{Supergiant Fast X--ray Transients: phenomenology and theory}

\begin{figure*}
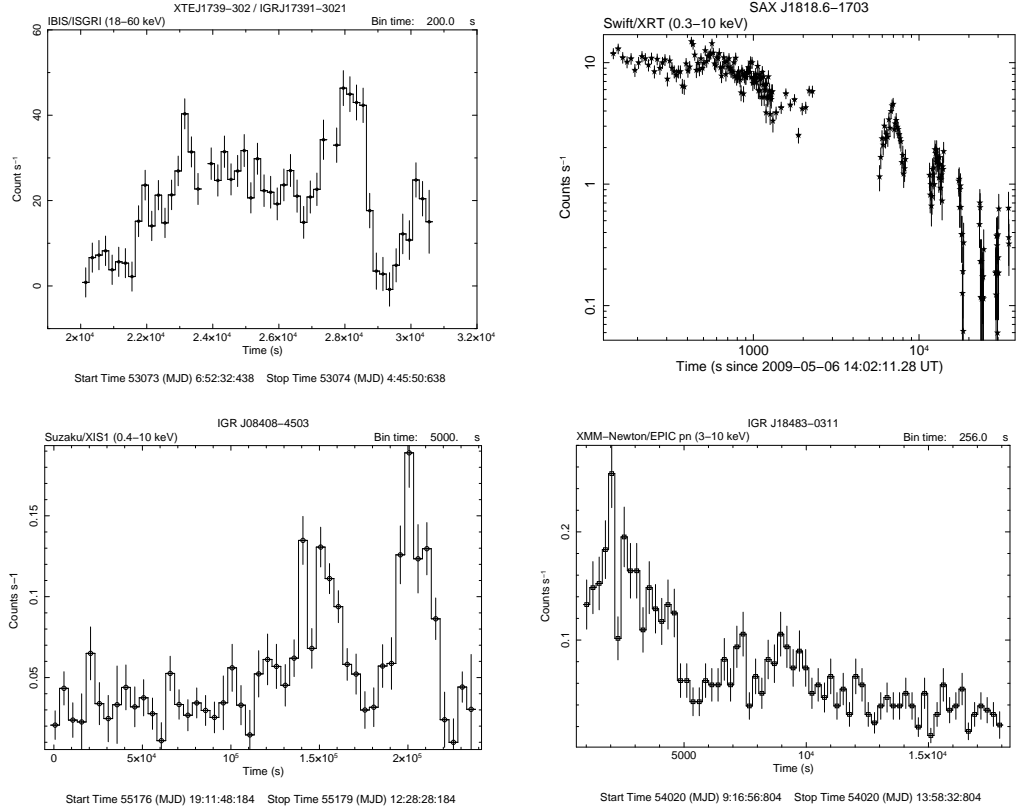

\centering
\begin{tabular}{cc}
\includegraphics[height=6.6cm, angle=-90]{lsfig2.ps} & \includegraphics[height=6.1cm, angle=-90]{lsfig3.ps} \\
\includegraphics[height=6.6cm, angle=-90]{lsfig4.ps} & \includegraphics[height=6.6cm, angle=-90]{lsfig5.ps}
\end{tabular}
\caption{\scriptsize Four examples of light curves of SFXTs during their 
outbursts (\emph{Top panels}; L$_{X}$$\sim$10$^{36}$~erg~s$^{-1}$) and in a low  intensity state (\emph{Bottom panels}; 
L$_{X}$$\sim$10$^{33}$~erg~s$^{-1}$).
 }
\label{lsfig:lcurves}
\end{figure*}

The \inte\ satellite opened a new era for High Mass X--ray Binaries (HMXBs) science,
discovering a subclass of hard transient sources, the Supergiant Fast X--ray Transients
(SFXTs; Sguera et al. 2005, 2006; Negueruela et al. 2006) during the survey of the 
Galactic plane (Bird et al. 2010).
SFXTs display bright X--ray activity concentrated 
in short (from  a few minutes to a few hours)
flares which reach an X--ray luminosity of 10$^{36}$--10$^{37}$ erg s$^{-1}$, with 
a hard spectrum below 10 keV (photon index $\Gamma$$\sim$0--1; Walter et al. 2006, Sidoli et al. 2006). 
The flares are part of an outburst phase lasting 
a few days (Romano et al. 2007, 
Sidoli et al. 2009a). 
Outside outbursts, SFXTs are usually found 
in an intermediate luminosity state of 10$^{33}$--10$^{34}$~erg~s$^{-1}$
with a softer spectrum well modelled with 
an absorbed power law with $\Gamma$$\sim$1--2  (Sidoli et al. 2008).
This hard emission and the faint flux variability 
(Fig.~\ref{lsfig:lcurves}, {\em Bottom panels}) suggest
a residual accretion onto the compact object (Sidoli et al. 2008, 2010; Bozzo et al. 2010).
The lowest luminosity state at 10$^{32}$~erg~s$^{-1}$ with a  very soft, thermal spectrum, 
has been rarely observed  (in't Zand et al. 2005).
SFXTs dynamic range (ratio of the peak luminosity in outburst to the  quiescent emission),
can span 3--5 orders of magnitudes.
The duty cycles are small, although highly variable from source to source: 
they are in the range 3\%--5\%, as derived from Swift/XRT monitoring of three SFXTs  (Romano et al. 2011),
while much smaller at hard X--rays as observed with \inte\ (Ducci et al. 2010, Sidoli 2010).

The optical counterparts are OB supergiants
(e.g. Masetti et al. 2006; Pellizza et al. 2006; Nespoli et al. 2008; Rahoui et al. 2008; Ratti et al. 2010),
implying that SFXTs are HMXBs similar to the supergiant HMXBs (SGXBs)
persistently emitting at X--rays.
Indeed, the main open issue is the mechanism which determines
the fast transient (SFXTs) instead of the persistent (SGXBs) X--ray emission (see below).
The discovery of these transients, only sporadically bright in X--rays, 
is important because they might be a dominant population 
of hidden HMXBs, which are good tracers of the recent
star formation rate (Mineo et al. 2010). 
So they are  not only interesting for  the investigation
of the accretion mechanisms,
but also for the study of the massive star formation,  
the chemical enrichment of our Galaxy and the evolutionary path of massive binaries.
The class includes 10 members to date (Table~\ref{lstab:sfxt}), 
with several  candidates  where an optical identification is uncertain or still missing. 
A recently proposed candidate is IGR~J16328--4726 (Fiocchi et al. 2010). 
Remarkably, these numbers are comparable to all SGXBs discovered since the birth of X--ray astrophysics,
before the advent of \inte.

%%%%%%%%%%%%%%%%%%%%%%%%%%%%%%%%%%%%%%%%%%%%%%%%%%
\begin{figure}[ht!]
\begin{center}
\vspace{-0.5truecm}
\includegraphics*[angle=270,scale=0.45]{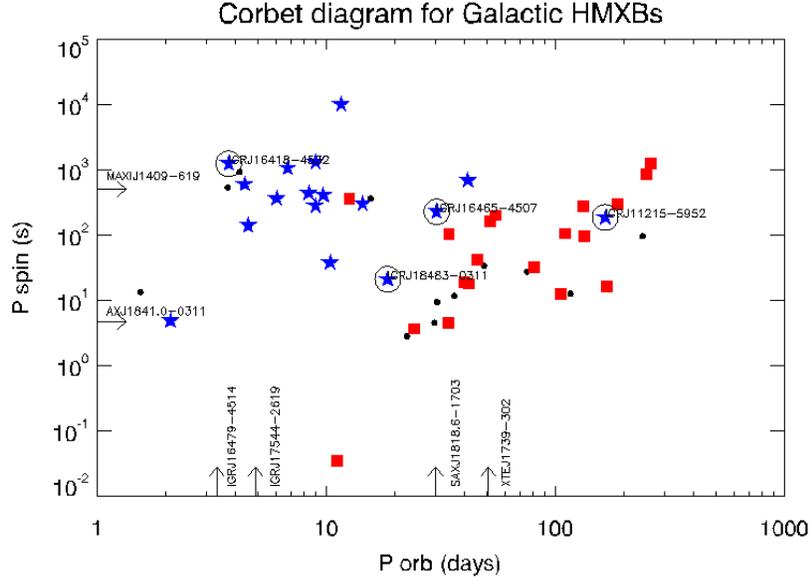}
\end{center}
\vspace{-0.75truecm}
\caption{\scriptsize Corbet diagram of Galactic HMXBs  (updated to February 2011).
{\em  Blue stars} mark HMXBs with optically identified supergiant companions; {\em  red squares} mark massive binaries
with optically identified Be donors (Liu et al., 2006).
{\em Small black dots} indicate HMXBs where the spectral type of the companion star is still uncertain.
{\em Large circles} around blue stars mark Supergiant Fast X--ray Transients. 
Arrows mark the position of SFXTs (or candidates SFXTs) where only
orbital or spin period is known to date.
}
\label{lsfig:corbet}
\end{figure}
%%%%%%%%%%%%%%%%%%%%%%%%%%%%%%%%%%%%%%%%%%%%%%%%%%

%%%%%%%%%%%%%%%%%%%%%%%%%%%%%%%%%%%%%%%%%%%%%%%%%%%%%%%%
%%%%%%%%%%%%%%%%%%%%%%%%%%%%%%%%%%%%%%%%%%%%%%%%%%%%%%%%

%%%%%%%%%%%%%%%%%%%%%%%%%%%%%%%%%%%%%%%%%%%%%%%% TABLE 1
\begin{table}
\vspace{-0.2truecm}
\caption{List of firm Supergiant Fast X--ray Transients with their  orbital and spin periodicities. }
\begin{center}
\begin{small}
\begin{tabular}{lccc}
\hline
\hline
Source           & Orbital Period (days)  &  Spin Period (s)   & Ref~$^a$     \\
\hline
IGR~J08408--4503 & 35 (?)               & -                    & [1]  \\
IGR~J11215--5952 & 164.6                & 186.78$\pm{0.3}$     & [2,3,4,5] \\
IGR~J16418--4532 & 3.753$\pm{0.004}$    & 1246$\pm{100}$       & [6,7]         \\  
IGR~J16465--4507 & 30.243$\pm{0.035}$   & 228$\pm{6}$          & [8,9,10]  \\
IGR~J16479--4514 & 3.3194$\pm{0.0010}$  & -                    & [11]  \\
XTE~J1739--302   & 51.47$\pm{0.02}$     & -                    & [12] \\
IGR~J17544--2619 & 4.926$\pm{0.001}$    & -                    & [13]  \\
SAX~J1818.6--1703& 30.0$\pm{0.1}$       & -                    & [14, 15] \\
AX~J1841.0--0536 & -                & 4.7394$\pm{0.0008}$  &    [16]  \\
IGR~J18483--0311 & 18.55 $\pm{0.03}$    & 21.0526$\pm{0.0005}$ &  [17,18]   \\
\hline
\end{tabular}
\end{small}
\end{center}
\begin{scriptsize}
{$^a$~The numbers correspond to the following references: [1]-Romano et al. 2009a; [2,3]-Sidoli et al. 2006, 2007; [4]-Romano et al. 2009b; 
[5]-Swank et al. 2007; [6]-Corbet et al. 2006; [7]-Walter et al. 2006; [8]-Clark et al. 2009; [9]-La Parola et al. 2010; [10]-Lutovinov et al. 2005;
[11]-Jain et al. 2009; [12]-Drave et al. 2010; [13]-Clark et al. 2009; [14]-Zurita Heras \& Chaty 2009; [15]-Bird et al., 2009; [16]-Bamba et al., 2001;
[17]-Levine et al. 2006, [18]-Sguera et al. 2007.}
\end{scriptsize}
\label{lstab:sfxt}
\end{table}
%%%%%%%%%%%%%%%%%%%%%%%%%%%%%%%%%%%%%%%%%%%%%%%%%%%%%%%%%%%%%%%%%%%%%%%%%%%%%%%%%%%%%%%%%%

A broad-band spectroscopy (0.3--100 keV) 
is possible only during the brightest flares, to date. 
This hampers an in-depth investigation
of the evolution of the high-energy part of the spectrum, i.e. the  
evolution of the Comptonizing medium. 
Phenomenological models, like  power-law with  exponential cut-off, have
been usually adopted, resulting
in hard photon indicis ($\Gamma$$\sim$0--1) and cut-off at 10--30 keV.
These spectra are similar to those displayed by accreting X--ray pulsars, 
suggesting that even in SFXTs where X--ray pulsations have not yet been found, a neutron star is present.
An in-depth spectroscopy with a more physical model 
has been performed for the SFXT with the lowest column density (10$^{21}$~cm$^{-2}$), 
IGR~J08408--4503, allowing us a deconvolution of the X--ray spectrum
with  a black-body together with a Comptonizing hot plasma (Sidoli et al. 2009c).
The fit resulted into two distinct photon populations, a
cold one (0.3~keV) likely coming from a thermal halo around the neutron star and a hotter
one (1.4--1.8~keV) probably originating from the accretion column onto the polar caps of the neutron star.
Sometimes SFXTs display absorption in excess of that towards the optical counterpart.
A variable column density has been observed in a few cases,
indicative of  local  absorbing matter (Romano et al. 2009a, Sidoli et al. 2009c), 
due to the clumpy nature of the supergiant wind (Rampy et al. 2009).

The discovery of X--ray pulsations demonstrates that at least in a half of the systems
the compact object is a neutron star (see Table~\ref{lstab:sfxt} and references therein). 
Orbital periods have also been measured, 
ranging from 3.3~days to 165~days (Table~\ref{lstab:sfxt}). 
Four SFXTs have both a measured spin and orbital period, thus it is possible to place
them in the Corbet diagram for Galactic HMXBs (see Fig.~\ref{lsfig:corbet}).
Interestingly, some SFXTs lie in the region of the diagram typical for Be transients (Fig.~\ref{lsfig:corbet}), 
possibly implying either a similar accretion mechanism (Sidoli et al. 2007) 
or an evolutionary link between these two classes of  HMXBs (Chaty 2010).
The shape of the bright flaring activity on short timescales from a few minutes to a few hours 
can be very complex  (see Fig.~\ref{lsfig:lcurves}, \emph{Top panels}): 
sometimes it is multi-peaked (IGR~J08408--4503, Romano et al. 2009a) 
with several re-flares (SAX~J1818.6--1703, Sidoli et al. 2009b), 
sometimes with a quasi-periodic behavior  
(XTE~J1739--302, Smith et al. 1998; Ducci et al. 2010), 
or  with a fast rise and exponential decay evolution 
(IGR~J16479--4514, Ducci et al. 2010).
Each SFXT can exhibit different flare shapes, thus complicating the interpretation
of their phenomenology. 
A single outburst is composed by several  short flares, as is evident from
$Swift$/XRT monitoring of the outburst in IGR~J11215--5952 (Romano et al. 2007) or from the almost 
continuous $Suzaku$ observation
(lasting 65~hr) of a bright accretion phase in  IGR~J17544--2619 (Rampy et al. 2009).
Interestingly, here very different flare morphologies  coexist in a single  outburst: 
several short flares lasting $\sim$3~minutes and with a symmetric profile 
are present together with a very bright
flare compatible with an exponential  decay time of $\sim$30~minutes and reaching 
a peak about 10,000 times brighter than the first
10~hr of this observation. 
The flares are part
of an outburst lasting $\sim$2 days, showing a smooth increasing of the underlying
X--ray emission on  timescales of hours, 
with a double-peaked modulation, with a primary (brighter) and a secondary (fainter) peak, after  15~hrs
from the primary.
The orbital period of IGR~J17544--2619  is 4.926$\pm{0.001}$~days (Clark et al. 2009). 
Assuming their ephemeris, the $Suzaku$ observation covers the orbital phases from 
0.86 to 0.41 ($\phi$=0 in Clark et al. 2009 corresponds
to the maximum luminosity), confirming that outbursts are triggered at similar orbital
phases, likely corresponding to the periastron passage.
Note that the modulation in the folded light curve in IGR~J17544--2619 
is still present even excluding the outbursts (Clark et al. 2009).
In another prototypical SFXT, XTE~J1739--302, the periodic signal (51.47$\pm{0.02}$ days) appears to be generated {\em only} 
by the outburst emission (Drave et al. 2010). 
 
Usually these long-term periodicities are interpreted as the orbital
period of the binary systems (see Table~\ref{lstab:sfxt}) and have been derived
from a  temporal analysis on a large database (\inte/IBIS or \sw/BAT). 
The case of IGR~J11215--5952 is very different: this is the first SFXT where the
outbursts were discovered to occur strictly periodically (Sidoli et al. 2006)
and no bright X--ray emission have ever been observed outside this periodicity 
($\sim$165~days, Sidoli et al. 2007, Romano et al. 2009b). 
This is remarkable since it indicates that the outburst are {\em only} triggered near the periastron passage
and {\em only} during a very short orbital phase interval.
The brightest X--ray emission is too short to be produced only by the enhanced accretion onto the neutron star
approaching the periastron. 
We suggested that a stable supergiant wind structure is present along the orbit 
in the form of a preferential plane for the outflowing
wind (Sidoli et al. 2007), with the 
outbursts triggered near the periastron 
when the neutron star crosses this additional wind component, 
confined in a plane inclined with respect to the orbit.
The presence of supergiant outflow of this kind, reminiscent of the equatorial disks
present in Be stars, is not confirmed yet by observations at other wavelengths, although 
a very thin disk  cannot be ruled-out in
IGR~J11215--5952 by optical spectroscopy of the B0.5~Ia companion HD~306414  (Lorenzo et al. 2010).
Also in IGR~J18483--0311 (Sguera et al. 2007, Romano et al. 2010) and in the eclipsing 
IGRJ~16479--4514 (Bozzo et al. 2009) the bright outbursts take place
preferentially around a particular orbital phase.
Remarkably, all these SFXTs show very different orbital periods (see Table~1).
This might indicate that the orbital eccentricity and/or the possible 
presence of a stable structure in the supergiant wind
crossed by the compact object along the orbit, 
play an important role in triggering their transient activity.
This scenario explains also in a simple way the orbital light curve profile with three peaks 
observed in XTE~J1739--302 (see Drave et al. 2010 for details).

SFXTs are wind-accretors and the short X--ray flares composing the outbursts 
are thought to be produced 
by the accretion of the very dense wind clumps (in't Zand 2005; 
Walter \& Zurita Heras, 2007; Negueruela et al. 2008; Ducci et al. 2009). 
The compact object acts as a probe of the stellar wind properties,
since the accretion luminosity L$_{\rm X}$ is proportional to $\dot{M}_{w}$$v_{rel}$$^{-4}$,
where $\dot{M}_{w}$ is the wind mass loss rate, and  $v_{rel}$ is the relative velocity
between the compact object and the wind, which can be approximated, in long period binaries, 
with the wind terminal velocity (Waters et al. 1989).
Actually, the wind clumps can be more massive than that estimated from the accretion
luminosity, if the neutron star accretion radius (R$_{\rm acc}$) is smaller than the clump radius (R$_{\rm cl}$).
In this case only a fraction of the clump will be accreted.
This can happen in eccentric and/or long period orbits, when the neutron star is far away 
from the supergiant and the clumps have expanded significantly (see Ducci et al. 2009 for details).
Karino (2010), starting from the clumpy wind model by Ducci et al. (2009),
suggested that the parameter which controls the fast transient
versus persistent behavior is 
the value of the ratio $r$=R$_{\rm acc}$/R$_{\rm cl}$ along the neutron star orbit.
There are 3 cases: (a)-if $r$ is always smaller than 1 along the neutron star orbit,
the source is very faint or quiescent; (b)-if  $r$ is always larger than 1, a persistent source is expected;
(c)-if  the value $r$=1 is crossed along the orbit, the source should behave like a SFXT. 
This implies a narrow window for the allowed parameter region for SFXTs in 
the diagram of the orbital period versus binary eccentricity.
The comparison with the data does not confirm this scenario for all known sources, 
possibly suggesting that other mechanisms are sometimes at work, besides the accretion from clumpy winds.

Another proposed explanation is that in SFXTs the accretion is inhibited for most of the
time by the presence of a centrifugal or a magnetic barrier (Bozzo et al. 2008).
In this scenario, the X--ray flares  can be produced
by a mild variability in the wind density, 
if the neutron star has a very high  magnetic field (B$\sim$10$^{14}$--10$^{15}$~G) 
and a slow spin period  (P$_{spin}$ $\sim$1000~s).
However, magnetars in SFXTs have not been found yet.  
Instead, an indication for a low magnetic field  (B$\sim$10$^{11}$~G) 
has been obtained in a SFXT (Sguera et al. 2010).

\acknowledgments
\begin{small}
I would like to thank the organizers of the session ``P3: Transient phenomena'', Chryssa Kouveliotou 
and Jochen Greiner,
for their kind invitation at the 
25th Texas Symposium on Relativistic Astrophysics, held in Heidelberg on December 06-10, 2010.
This work was supported in Italy by ASI-INAF contract I/033/10/0 and by 
the grant from PRIN-INAF 2009, ``The transient X-ray sky: 
new classes of X--ray binaries containing neutron stars''
(PI: Sidoli).
\end{small}

\begin{scriptsize}

\end{scriptsize}

\end{document}